\documentclass[12pt,preprint]{aastex}

\newcommand{\mtip}{$M_I^{TRGB}$}
\newcommand{\itip}{$I^{TRGB}$}
\newcommand{\ome}{$\omega$ Centauri~}
\newcommand{\omen}{$\omega$ Centauri}

\slugcomment{to appear in Astrophysical Journal}

\shorttitle{The calibration of the RGB Tip}
\shortauthors{Bellazzini, Ferraro \& Pancino}

\begin{document}


\title{A step toward the calibration of the RGB Tip as a Standard Candle}

\author{Michele Bellazzini, Francesco R. Ferraro}
\affil{Osservatorio Astronomico di Bologna, Via Ranzani 1, 40127, Bologna, 
ITALY}
\email{bellazzini@bo.astro.it, ferraro@apache.bo.astro.it}

\and

\author{Elena Pancino\altaffilmark{1}} 
\affil{European Southern Observatory, K. Schwarzschild Str. 2, 
Garching, D-85748, Germany.}
\email{epancino@eso.org}

\altaffiltext{1}{on leave from Dip. di Astronomia,
Universit\`a di Bologna, Via Ranzani 1, 40127, Bologna, ITALY}

\begin{abstract}
The absolute $I$ magnitude of the Tip of the Red Giant Branch (\mtip)
is one of the most promising Standard Candles actually used in
astrophysics as a fundamental pillar for the Cosmological Distance
Scale. With the aim of improving the observational basis of its
calibration, we have obtained an accurate estimate of the (\mtip) for
the globular cluster \omen, based {\em (a)} on the largest photometric
database ever assembled for a globular, by \citet{panc}, and {\em (b)}
on a direct distance estimate for \omen, recently obtained by
\citet{ogle} from a detached eclipsing binary. The derived value \mtip
$=-4.04\pm 0.12$ provides, at present, the most accurate empirical
zero-point for the calibration of the \mtip -- [Fe/H] relation, at
[Fe/H]$\sim -1.7$. We also derived a new empirical \mtip$-[Fe/H]$
relation, based on the large IR dataset of red giants in Galactic
Globular Clusters recently presented by \citet{fer00}. This database
(extending up to $[Fe/H]=-0.2$) covers a more appropriate metallicity
range, for extragalactic applications, than previous empirical
calibrations (limited to $[Fe/H]\le -0.7$). The proposed relation is
in excellent agreement with the newly determined
zero-point.\thanks{Based on data taken at the European Southern
Observatory, Chile, as part of the ESO observing programmes 62.L-0354,
63.L-0349 and 65.L-0463.}

\end{abstract}


\section{Introduction}

While the use of the luminosity of the Tip of the Red Giant Branch
(TRGB) as a standard candle dates back to 1930 (see \citet{mf98} and
references therein), the development of the method as a safe and
viable technique is relatively recent (\citet{lfm93}, hereafter
L93). In a few years it has become a widely adopted technique, finding
fruitful application also within the {\em HST Key project on the
Extragalactic Distance Scale} (see \citet{ferr-a,ferr-b}, and
references therein). The underlying physical processes are clearly
understood and well rooted in the theory of stellar evolution
\citep{mf98}. The method is particularly useful to estimate distances
to those stellar systems that do not contain Cepheids, such as early
type galaxies, and it can be applied to galaxies as far as $\sim 12$
Mpc with the current instrumentation. The key observable quantity is
the magnitude of the bright end (the tip) of the Red Giant Branch
(RGB), that corresponds to a sharp cut-off in the RGB Luminosity
Function (LF), measured in the Cousin's $I$
passband\footnote{Hereafter, when we will refer to the $I$ passband,
we will always mean the {\em Cousin's} $I$ passband, unless we specify
otherwise.}. In this passband the magnitude of the tip shows a very
weak (if any) dependence on metallicity (\citet{da90}, hereafter
DA90). The feature can be identified by applying the Sobel's filter,
an edge-detection algorithm, to the LF of the upper RGB (see
\cite{smf96}, hereafter S96, for a standard application\footnote{From
now on, we will refer to their method for the detection of the TRGB as
to the {\em standard analysis}.}). The possible biases have been well
characterized and quantified by means of numerical simulations by
\citet{mf95} (hereafter MF95).

In this context, we are performing an extensive study of the Red Giant
population in Galactic Globular Clusters (GGC). In particular, we have
derived an accurate calibration of the RGB photometric properties as a
function of metallicity, both in the optical and in the IR
\citep{fer99,fer00}. The final aim of this program is to use stellar
populations in GGCs as {\it calibrators} for the TRGB method, taking
advantage of the large samples that can be easily assembled with the
new generation of array detectors and cameras.  As a part of this
general project, in this paper we present {\em (i)} a very robust
empirical calibration of \mtip at $[Fe/H]\sim -1.7$, derived by
applying the {\em standard analysis} to a very large sample of RGB
stars in the globular cluster \ome (NGC 5139), and based on a direct
distance estimate recently obtained by \citet{ogle} (hereafter T01)
for a detached eclipsing binary in this cluster, and {\em (ii)} a new
{\em empirical} calibration of the \mtip -- $[Fe/H]$ relation based on
a large, homogeneous IR database of RGB stars in Galactic Globular
Clusters, recently published by \citet{fer00} (hereafter F00).

The aim of this paper is to provide a first, robust pillar for a well
rooted empirical calibration of \mtip as a standard candle, extending
over a wide range in metallicity.  This will constitute the basis for
a secure application of the TRGB method as an extragalactic distance
indicator.

\section{A new $M_{I}^{TRGB}-[Fe/H]$ relation}

The most widely adopted calibration of the absolute I magnitude of the
TRGB (\mtip $=M_{bol}^{TRGB}-BC_I$) was derived by L93 from the
following relations, which give the bolometric magnitude of the tip
($M_{bol}^{TRGB}$) as a function of metallicity:

\begin{equation}
M_{bol}^{TRGB}=-0.19[Fe/H]-3.81
\end{equation}

and the I bolometric correction ($BC_I$) as a function of the intrinsic 
color $(V-I)_0$:

\begin{equation}
BC_I=0.881-0.243(V-I)_0
\end{equation}

both obtained by DA90. Using accurate V and I photometry of a small
number of RGB stars\footnote{From 20 to 110 stars per cluster were
observed by DA90 in the upper $\sim 4$ mag of each RGB.} in eight
selected globular clusters in the metallicity range $-2.2\le
[Fe/H]\le -0.7$, DA90 estimated the apparent I magnitude of the TRGB
(\itip) as the magnitude of the brightest and reddest giant in their
samples. They converted \itip to absolute magnitudes by adopting the
RR Lyrae distance scale $M_V(RRLy)=0.17[Fe/H]+0.82$, from \citet{ldz},
and finally to $M_{bol}$ using Eq. 2. Note that, in this context, the
most relevant observational fact presented by DA90 is that \mtip
$=-4.0 \pm 0.1$, in the sampled range of metallicity.

The above calibrations suffer from two major sources of uncertainty:
\begin{itemize}

\item{} The evolution of stars along the RGB becomes faster as their
luminosity increases along the path toward the TRGB. Thus, most of the
cluster light has to be observed to correctly sample the fastest
evolutionary phases \citep{rf88}. MF95 stated that acceptable
detections of the TRGB can be obtained if more than 50 stars are
sampled within 1 mag from the tip and that fine estimates can be
obtained only sampling more than 100 stars in that range. The DA90
samples are much poorer than this, therefore their \itip estimates may
be affected by the systematics associated with small number
statistics.

\item{} The RR Lyrae distance scale is quite uncertain. While there is now some
agreement on the slope of the $M_V(RRLy)$ -- $[Fe/H]$ relation, the
actual zero points is still hardly debated (see \citet{carla}, for a
recent review).
   
\end{itemize}

The most recent calibration of the $M_{bol}^{TRGB}-[Fe/H]$ relation
has been obtained by F00, from homogeneous near-infrared photometry of
a large sample of RGB stars in nine Galactic Globular
Clusters. Several hundred giants were observed in the upper $\sim 4$
mag of each RGB. Moreover, the data-set presented by F00 extends to a
much higher metallicity with respect to DA90, covering the range
$-2.2\le [Fe/H]\le -0.2$. The TRGB was identified by using the
brightest non-variable RGB star in K band. F00 adopted the new
distance scale presented by \citet{fer99} which, although still based
on the Horizontal Branch, was obtained by using a new semi-empirical
approach (see \citet{fer99} for details). F00 provided
$M_{bol}^{TRGB}$ as a function of [Fe/H] in two metallicity scales:
{\it (i)} the \citet{cg97} scale and {\it (ii)} the {\it global}
metallicity ([M/H]) scale, which takes into account also the
$\alpha-$elements abundances. To obtain a more direct comparison with
the calibration by L93, we used the data from F00 to derive
$M_{bol}^{TRGB}$ as a function of [Fe/H] in the \citet{zw84}
metallicity scale, and obtained:

\begin{equation}
M_{bol}^{TRGB}=-0.12[Fe/H]-3.76 .
\end{equation}
 
The larger samples used by F00, the wider metallicity range covered
and the minor impact of the extinction and bolometric corrections on
infrared luminosities, suggest that eq. 3 has to be preferred to
eq. 1.  Note, however, that the final \mtip -- [Fe/H] relations
derived by the two independent calibrations of $M_{bol}^{TRGB}$ are in
agreement within the uncertainties (see Sect. 4 and Fig. 4).

In order to derive the \mtip\ from the new empirical calibration, we
need to combine eq. 3 with eq. 2, following the relation
$M_{I}^{TRGB}=M_{bol}^{TRGB}-BC_I$. The actual $BC_I$ depends
implicitly on metallicity, since $BC_I$ depends on the $(V-I)$ color
of the TRGB (eq. 2) which, in turn, depends on metallicity. To make
such a dependence explicit, we plot in Fig. 1 $(V-I)_0^{TRGB}$ as a
function of [Fe/H] for the six clusters studied by DA90. The data are
best fitted by a second order polynomial whose equation is also
reported in Fig. 1.  We adopt this relation to eliminate the
dependency of \mtip\ on $(V-I)$ color and we finally obtain the
following relation:
 
\begin{equation}
M_{I}^{TRGB}=0.14[Fe/H]^{2}+0.48[Fe/H]-3.66 
\end{equation}
 
which provides the calibration of \mtip\ as a function of the 
metallicity only (in the \citet{zw84} scale). 

Eq. (3) and (4) represent a substantial improvement with respect to
previous work, since they are based on the largest IR RGB samples in GGCs
available in the literature.  Note that the F00 survey sampled a
significative fraction of the cluster light (up to $\sim 80\%$).
However even in this large data-set the number of RGB stars in the
upper 1 mag bin from the TRGB is $< 50$ in all the cases, and the
brightest star detection is possibly prone to low number statistics
effects.  This is due, in most cases, to intrinsic poorness of the
cluster population, since the majority of GGCs simply {\em do not
contain a sufficient number of RGB stars} to adequately sample the
upper RGB. Thus extensive observations of the most massive GGCs are
urged in order to properly calibrate the above relations.  In the
following sections we report our results on the most massive Galactic
globular: $\omega$ Centauri.

\begin{figure}
\epsscale{1.0}
\plotone{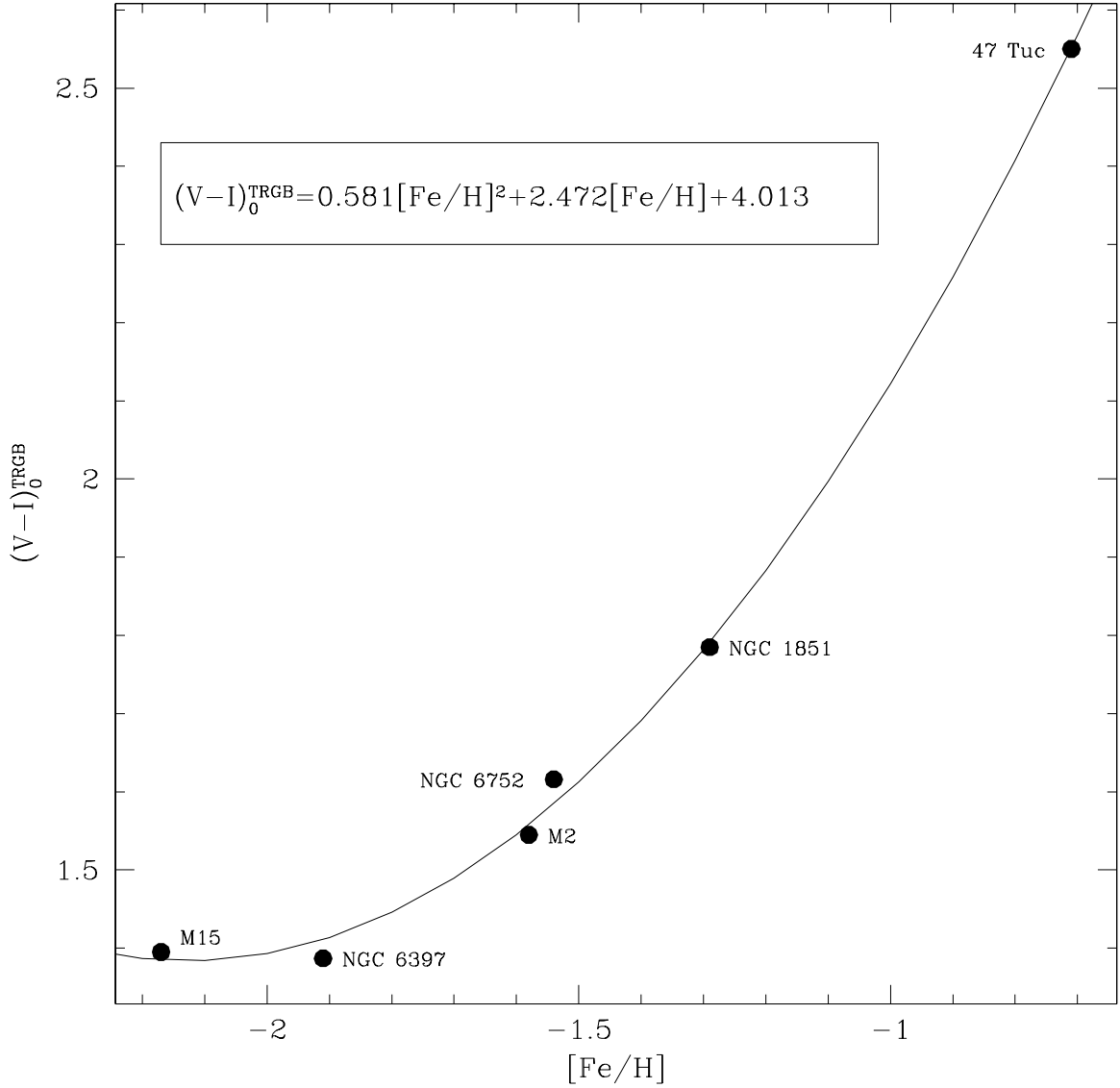}
\caption{$(V-I)_0^{TRGB}$ as a function of [Fe/H] for the six clusters studied 
by DA90. All data are taken from DA90. The curve represents the best
fits to the data, the corresponding equation is also reported. 
The $rms$ scatter is $0.04$ mag.}
\end{figure}

\section{The determination of the TRGB luminosity  in \ome}

\subsection{Why \ome?}

\ome is a nearby and well studied cluster. It is the most luminous  
globular cluster in the Milky Way system, therefore even the fastest
evolutionary phases are well populated and it is possible to observe a
quite large sample of RGB stars, fulfilling the prescriptions of
MF95. Here we adopt the huge photometric ($B$,$I$) database presented
by P00, consisting of more than 220,000 stars observed with the WFI
camera at the 2.2 ESO-MPI telescope and extending over a $\sim
34^{\prime} \times 33^{\prime}$ field roughly centered on the cluster.
The absolute photometric calibration is accurate to within $\pm 0.02$
mag.  The observed field extends over $\sim 24 $ core radii
\citep{vanleuw}, i.e. an area enclosing $90 \%$ of the cluster light.
Thus, virtually {\em all the bright stars of \ome are included in the
adopted database.} Considering that \ome is the most luminous globular
cluster of the whole Galaxy, this photometric database is the largest
sample that can be obtained in the Galactic globular cluster system.
 
Moreover, \ome was the first globular in which a detached eclipsing
binary system, {\it OGLEGC17}, was discovered \citep{kal96}.  This
enabled T01 to obtain a direct distance estimate, independent of any
other distance scale. The distance of detached, eclipsing double line
spectroscopic binaries can in fact be obtained by coupling the
physical parameters of the system with the relations between color and
surface brightness by \citet{ba76}. This method is basically {\em
geometrical}, since the distance is obtained by comparison between the
linear and angular size of the binary members (see
\citet{lacy,kr99}). The final distance obtained by T01 is $d= 5385\pm
300$ pc, which corresponds to $(m-M)_0=13.65\pm 0.11$, if we assume
their adopted extinction value, $E(B-V)=0.13\pm 0.02$. Their result is
in good agreement with previous estimates. According to T01, the error
bar on the distance modulus can be significantly reduced as soon as
better light and velocity curves will be obtained for {\it
OGLEGC17}. Thus, a significant improvement of the quoted measure of
the distance modulus has to be expected in the near future.  A further
independent distance estimate based on a geometrical method could soon
be provided also by the comparison of the linear and the angular velocity
dispersion from the large database of proper motions by \citet{vanleuw}, 
once supported by detailed dynamical modeling.

In deriving \mtip another fundamental ingredient can still be greatly
refined and improved in the near future: the {\em reddening}. For
instance, \citet{muskkk} report that the $(V-I)_0$ color at minimum
light of RRab variables is universal and independent of
metallicity. Thus, reliable $E(V-I)$ estimates can be obtained from
the {\em observed} $(V-I)$ at minimum light and large samples would
greatly improve the accuracy. For \omen, in particular, a ($V$,$I$)
survey of RR Lyrae can potentially provide a reddening estimate with an
uncertainty of $\pm 0.01$ mag or lower, since this populous cluster
contains more than 70 RRab \citep{rey}.

In conclusion, \ome seems to fit all the requirements to obtain an
excellent calibration of \mtip and still leaves considerable room for
future improvements.

A possible caveat may be associated with the wide metallicity
distribution observed in this cluster (see P00 and references
therein). However, it has been shown that the largely dominant
population is metal poor [\citet{n96,sk96}, P00]. The peak of such
population occurs at $[Fe/H]\sim -1.7$ [\citet{sk96}, P00] and we will
adopt this as the characteristic metallicity of the cluster.

\subsection{\mtip in \ome}

Fig. 2 shows the ($I$,$B-I$) Color Magnitude Diagram (CMD) of the
brightest 3 mag in the RGB of \ome: 1777 stars are reported in this
diagram, most of them being RGB stars. The contribution of AGB stars
is negligible for $I < 11$. The photometric errors are lower than 0.02
mag in both passbands, over the whole magnitude range  plotted
in the diagram. The contamination by foreground sources is negligible in this
region of the CMD and the crowding is not a serious concern for such
bright stars in this nearby and relatively loose cluster. 

The P00 sample contains 185 stars in the upper magnitude bin, which is
almost twice of what recommended by MF95 for an optimal detection of
the TRGB.

\begin{figure}
\epsscale{1.0}
\plotone{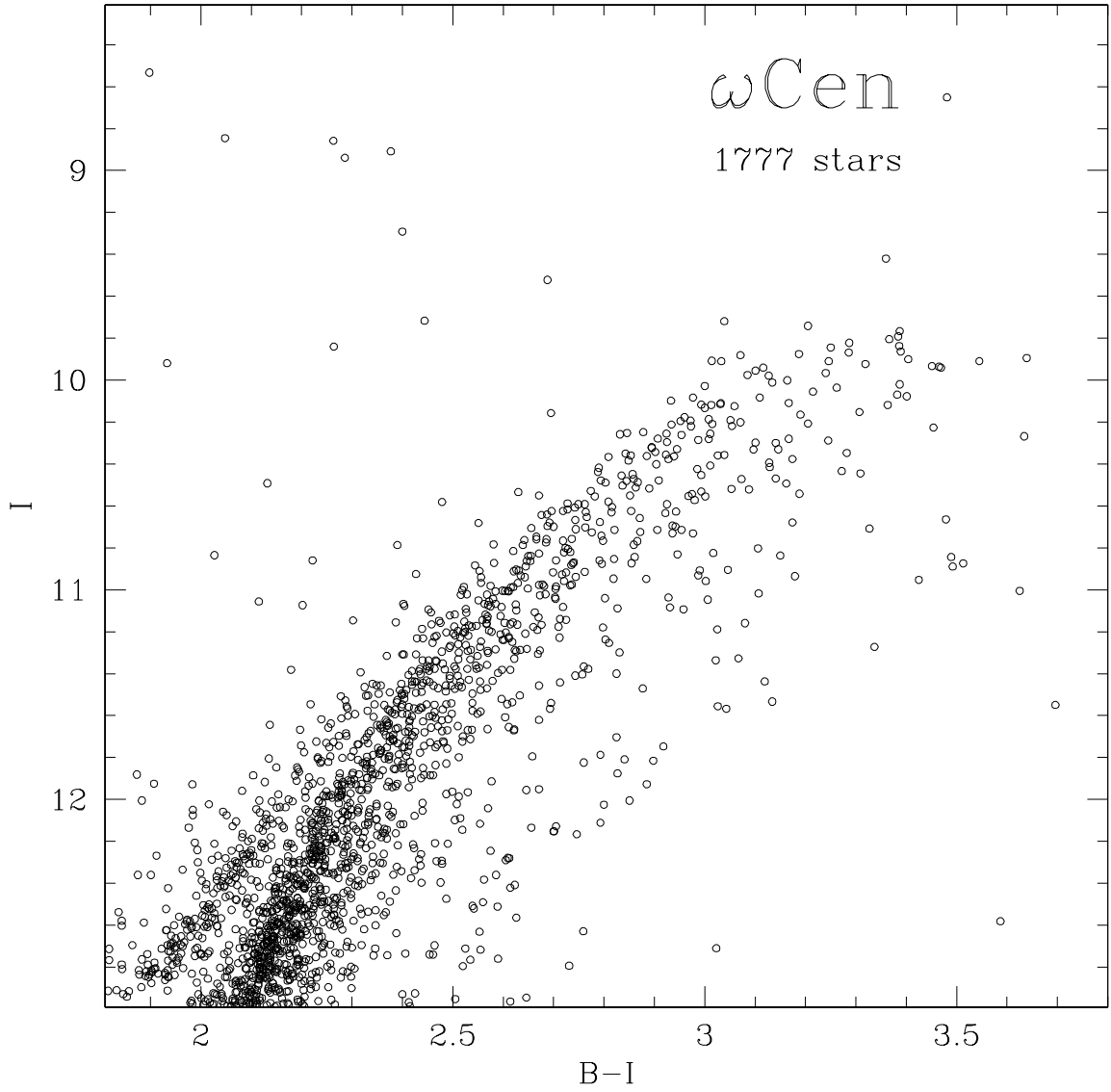}
\caption{($I$,$B-I$) CMD of the brightest 3 mag of the RGB of \omen, 
from P00}
\end{figure} 

The edge-detector filter is applied on a smoothed version of the LF of
the RGB, following the {\em standard analysis} as described by SMF96
(see their appendix). In Fig. 2 {\em panel (a)} the RGB LF for $I<11$
is reported.  A sharp cut-off is clearly evident at $I\sim 9.8$. The
same LF is shown in Fig. 2 {\em panel (b)} as a smoothed histogram,
while the edge-detector filter response to the smoothed LF is reported
in {\em panel (c)}.  The main peak in the filter response indicates
the cut-off point: the TRGB is clearly and unambiguously detected at
$I=9.84\pm 0.1$, the associated uncertainty is the Half Width at Half
Maximum (HWHM) of the peak.

\begin{figure}
\epsscale{1.0}
\plotone{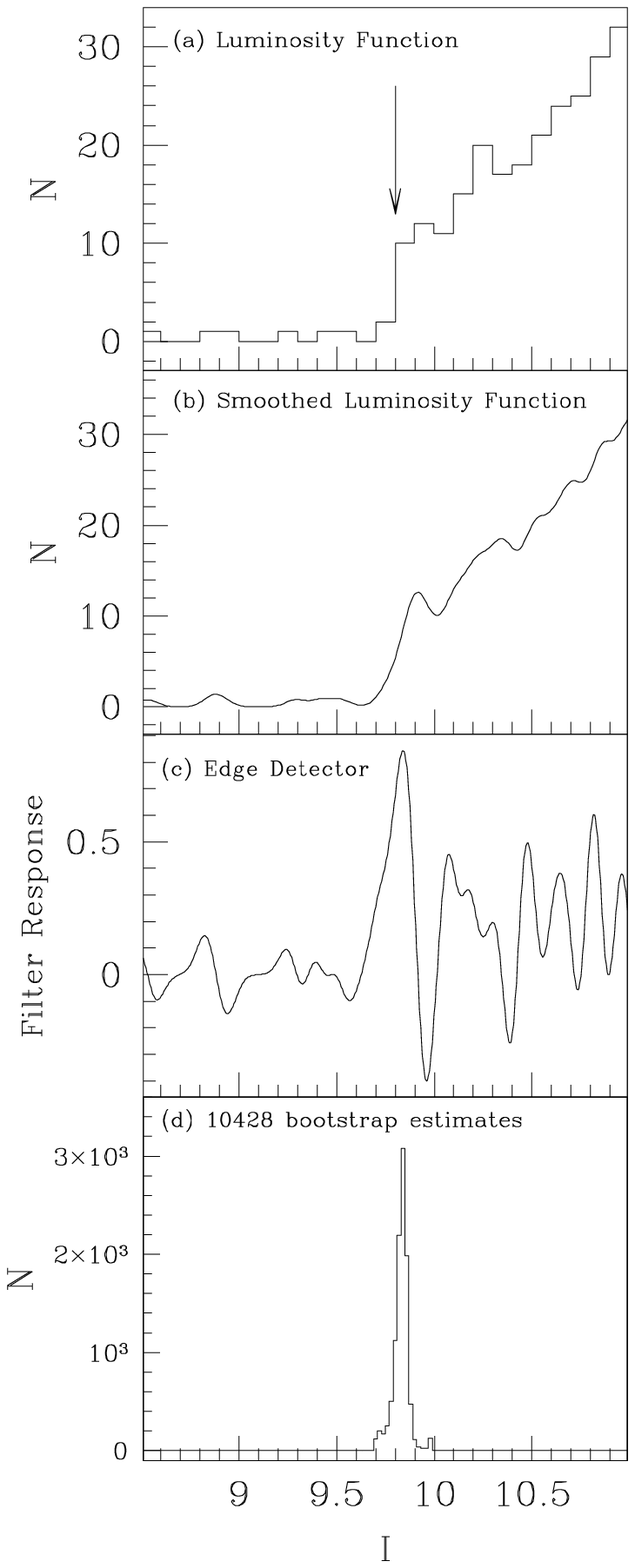}
\caption{The LF for the RGB stars with $I<11$ is shown as a histogram 
in {\em panel (a)} and as a smoothed histogram in {\em panel (b)}. 
The arrow in {\em panel (a)} indicates the cut off of the LF corresponding to 
the TRGB.
The smoothed histogram is multiplied by an arbitrary constant to facilitate
comparison with {\em panel (a)}. The response of the edge-detection
algorithm is shown in {\em panel (c)}, while the distribution of the
bootstrapped estimates is shown in {\em panel (d)}. The location of \itip\
is unambiguously identified by the peaks in {\em panels (c)} and {\em (d)} at
$I=9.84\pm 0.1$.The associated uncertainty is the Half Width at Half Maximum 
(HWHM) of the peak. }
\end{figure}

Moreover, the large sample of giants observed in this cluster allows
us to adopt a more refined approach to the quantification of the
uncertainty, by applying a {\it bootstrap} technique (see
\citet{astrostat}, and references therein). To do this, we randomly
extracted a subsample containing $80 \%$ of the stars from the global
sample of all RGB stars with $I<11$, we repeated the measure of \itip\
on the extracted subsample with the same technique as above and
recorded the final result. {\em Panel (d)} in Fig. 2 shows the
distribution of 10,428 such estimates on randomly extracted
subsamples: the mean of the distribution is $<I>=9.835$ and the
standard deviation is $\sigma=0.04$ mag. The estimate of \itip\ turns
out to be remarkably robust to statistical fluctuations, with $83 \%$
of the estimates falling within $\pm 1\sigma$ from the mean. We adopt
the bootstrapped $\sigma$ as the uncertainty on our estimate of the
apparent $I$ magnitude of the tip. The final result is: \itip $=9.84
\pm 0.04$.

It is now straightforward to provide \mtip\ of \ome as a function of
distance modulus and reddening:

\begin{equation}
M_{I}^{TRGB}=9.84 (\pm 0.04) - K \cdot E(B-V) - (m-M)_0
\end{equation}

where $K=A_I/E(B-V)$, $A_I$ is the amount of extinction in the $I$
passband, $E(B-V)$ is the reddening and $(m-M)_0$ is the true distance
modulus. Since the P00 data-set sampled virtually the whole cluster
population in this range of magnitudes, it is unlikely that the \itip\
estimate will be significantly improved by new observations. On the
other hand, all other terms of Eq. 5 can be subject to substantial
improvements in their estimates, as discussed in Sect. 3.1. Here we
adopt the distance and reddening by T01 and the reddening laws by
\citet{dean}\footnote{As far as we know \citet{dean} are the only ones
providing a reddening law for the Cousin's $I$ passband. All other
sources we were able to retrieve refer to Jhonson's $I$, a quite
different filter with different $A_I/E(B-V)$ values.}, with the same
assumptions of DA90, i.e., $A_I\simeq 1.76E(B-V)$. We obtain:

\begin{equation}
M_{I}^{TRGB}= -4.04 \pm 0.12
\end{equation}

where {\em all} the sources of uncertainty have been taken into
account. The main contributor to the error budget remains the estimate
of the distance modulus.

\section{Comparison with other calibrations}

It is worth checking how the existing empirical and theoretical
calibrations compare with the above estimate of \mtip, at the
relevant metallicity $[Fe/H]\sim-1.7$.

\begin{figure}
\epsscale{1.0}
\plotone{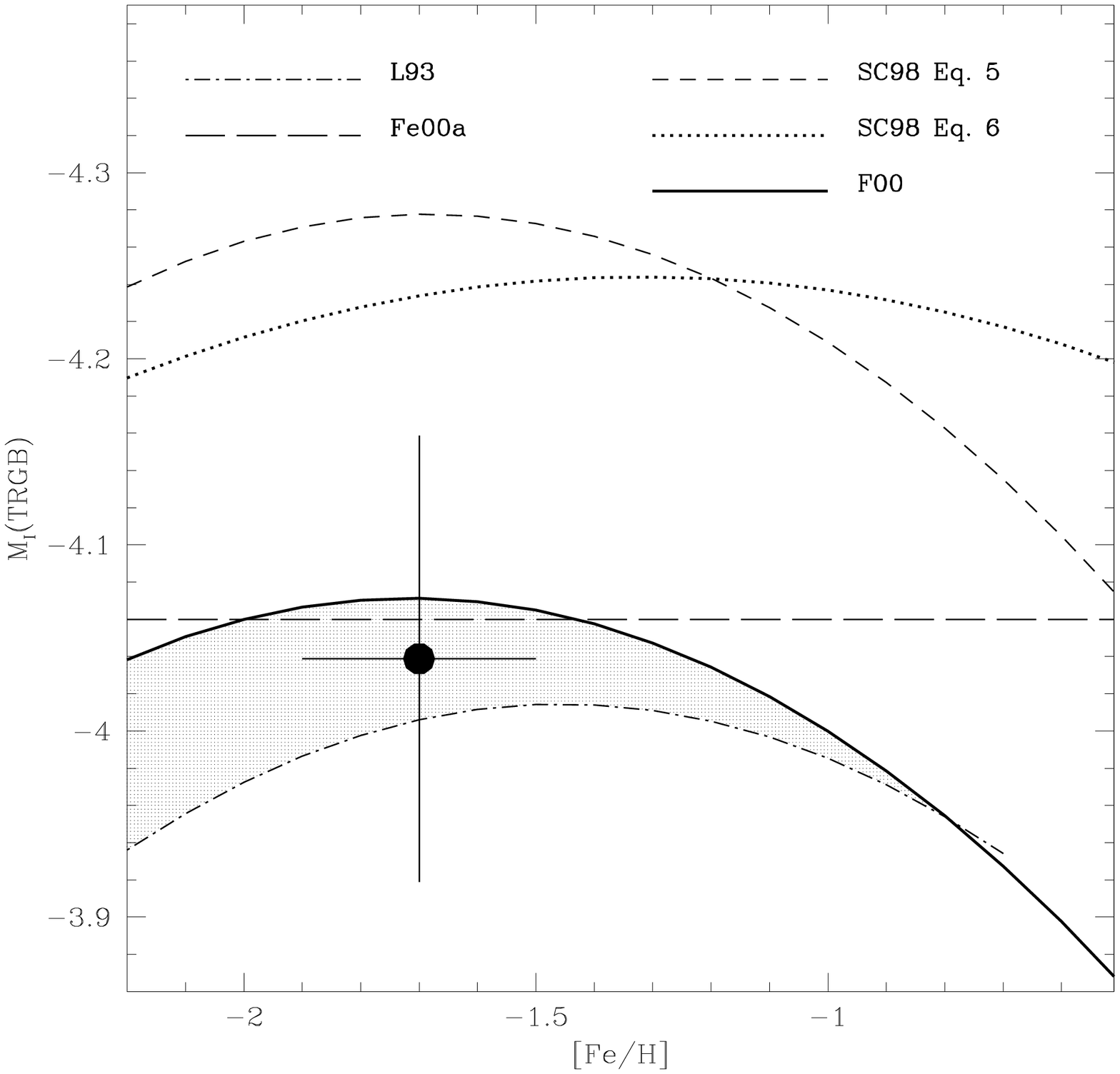}
\caption{Comparison between the calibration of \mtip\ obtained for \ome and
different \mtip\ -- [Fe/H] relations.  The {\em big black dot} is our
estimate of \mtip in \omen, where the horizontal error bar represents
the range in metallicity comprising the dominant population of the
cluster.  Different curves represent different calibrations: the {\em
short dashed curve} shows the calibration by SC98 (their eq. 5); the
{\em dotted curve} represents the same calibration with different
assumptions for the bolometric correction (eq. 6 in SC98); the {\em
dashed-dotted curve} is the empirical calibration by L93 (eq. 1 plus
eq. 2, reported in the \mtip\ -- [Fe/H] plane as described in
Sect. 2); the {\em heavy continuous curve} represents the empirical
calibration we have derived from the F00 database (our eq. 4); the
{\em long dashed line} reports the calibration of \mtip\ as a
secondary, indicator provided by Fe00a. The {\em shaded area} delimits
the region of the plane where most of the empirical relations
lie. Note that both the \ome measure and the Fe00a calibration fall
within this region.}
\end{figure}

In Figure 4 our estimate of \mtip\ in \ome is reported as a {\em big
black dot} in the \mtip vs. [Fe/H] plane. The horizontal error bar
represents the range in metallicity covered by the dominant population
of the cluster (see Fig.11 in \citet{sk96}).

The empirical calibrations discussed in Section 2 are also plotted:
the {\em heavy solid curve} is the new calibration based on the large
IR database by F00 (eq. 4), while the L93 calibration (eq. 1 plus
eq. 2 from DA90) has been converted to a function of metallicity,
following the procedure described in Sect. 2, and it is plotted as a
{\em dotted-dashed curve}. The horizontal {\em long dashed line}
represents the recent result by \citet{ferr-a} (hereafter Fe00a) who
calibrated the TRGB as a secondary indicator by using Cepheid
distances in a small set of nearby galaxies where both Cepheids and
the TRGB have been detected. They found \mtip $=-4.06\pm 0.07$
(random) $\pm 0.13$ (systematic), in very good agreement with our
estimate, despite their larger uncertainty. In Fig. 4, we plotted also
the region delimited by the two empirical calibrations discussed above
as a {\it shaded area}, in order to point out the region of the plane
where most of the empirical estimates lie.

\citet{sc98} (hereafter SC98) provided a theoretical relation of \mtip\ as a 
function of the global metallicity [M/H], instead of [Fe/H].  We
convert their [M/H] to [Fe/H] according to the prescriptions of
\citet{scs93} and assuming $[\alpha/Fe]=+0.28$ over the considered
metallicity range (see \citet{fer99}). Using two different bolometric
corrections, SC98 derived two slightly different relations (their
eq. 5 and 6), which are reported in Fig. 4 as a {\em short dashed
curve} and as a {\em dotted curve}, respectively.

As can be seen from Fig. 4, the {\em theoretical} calibrations are
systematically ($\sim 0.2$ mag) brighter than {\em empirical} ones, as
already noted by SC98 and Fe00a. The strong constraint provided by the
TRGB luminosity in \ome seems to favor the empirical calibrations.
However, as stated by \citet{cdl} and F00, theoretical relationships
should be considered as upper limits for the TRGB luminosity, since
they refer to the nominal red giant luminosity at the He flash.

\section{Summary and Conclusions}

We have used some of the most recent results from the study of
Galactic Globular Clusters to improve the observational basis of the
calibration of the TRGB method as a powerful distance indicator.

In particular, we have derived a robust estimate of \mtip in the
globular cluster \omen, based {\em (a)} on the application of the
standard technique on the very large sample of RGB stars by P00, and
{\em (b)} on a direct distance estimate from a detached eclipsing
binary by T01.  Our result turns out to be in excellent agreement with
previous empirical calibrations.  Expected improvements in the
distance and reddening estimates may likely reduce the uncertainties
of our calibrating point to $\sim \pm 0.06$ mag. Thus, eq. 5 shall be
considered as a very firm point of the TRGB calibration (at
$[Fe/H]\sim -1.7$) for the future, as well as an important
observational test for stellar evolutionary models. However, even at
the present level of accuracy, the \mtip measure derived here is the
less uncertain calibrating point for the \mtip -- [Fe/H] relation.

A new empirical \mtip-$[Fe/H]$ relation (eq. 4) was also derived.  It
turns out to be in excellent agreement with the derived \mtip of \ome
as well as with other empirical calibrations.  The main advantages of
the new relation derived here with respect to the previous ones (L93,
DA90) are: {\it (i)} it is based on more solid observational basis,
since it has been derived from the largest IR database of RGB stars in
Galactic globulars available in the literature (F00), {\it (ii)} it
has been calibrated over a larger range of metal abundances, extending
the metallicity limit from $[Fe/H] \le -0.7$ (L93, DA90) to $[Fe/H]
\le -0.2$; for this reason, eq. 4 has to be considered a much more
appropriate tool for applications to external galaxies, which, in
general, are significantly more metal rich than $[Fe/H] > -0.7$.
 
\acknowledgments
This research has been supported by the Italian Ministero della
Universit\`a e della Ricerca Scientifica e Tecnologica (MURST),
through the COFIN grant p. MM02241491\_004, assigned to the project
{\em Stellar Observables of Cosmological Relevance}.  The financial
support of the Agenzia Spaziale Italiana (ASI) is also kindly
acknowledged.  We thanks M. Tosi, G. Clementini and L. Greggio for
useful discussions. E.P. acknowledges the ESO Studentship
programme. This research has made use of NASA's Astrophysics Data
System Abstract Service.

\end{document}